# Routing Security Issues in Wireless Sensor Networks: Attacks and Defenses

Jaydip Sen
*Innovation Lab, Tata Consultancy Services Ltd.*
*India*

## 1. Introduction

Wireless Sensor Networks (WSNs) are rapidly emerging as an important new area in wireless and mobile computing research. Applications of WSNs are numerous and growing, and range from indoor deployment scenarios in the home and office to outdoor deployment scenarios in adversary's territory in a tactical battleground (Akyildiz et al., 2002). For military environment, dispersal of WSNs into an adversary's territory enables the detection and tracking of enemy soldiers and vehicles. For home/office environments, indoor sensor networks offer the ability to monitor the health of the elderly and to detect intruders via a wireless home security system. In each of these scenarios, lives and livelihoods may depend on the timeliness and correctness of the sensor data obtained from dispersed sensor nodes. As a result, such WSNs must be secured to prevent an intruder from obstructing the delivery of correct sensor data and from forging sensor data. To address the latter problem, end-to-end data integrity checksums and post-processing of senor data can be used to identify forged sensor data (Estrin et al., 1999; Hu et al., 2003a; Ye et al., 2004).

The design and implementation of secure WSNs must simultaneously address several difficult research challenges. First, wireless communication among the sensor nodes increases the vulnerability of the network to eavesdropping, unauthorized access, spoofing, replay, and *denial-of-service* (DoS) attacks. Second, the sensor nodes themselves are highly resource-constrained in terms of limited memory, CPU, communication bandwidth, and especially battery life. These resource constraints limit the degree of encryption, decryption, and authentication that can be implemented on individual sensor nodes, and call into question the suitability of traditional security mechanisms such as computation-intensive public-key cryptography for such resource-constrained sensor nodes (Carman et al., 2000). Third, WSNs face the added physical security risk of individual sensor nodes falling into wrong hands. Sensor nodes that are physically deployed in the field can be captured by an intruder, and can then be subject to attacks from the potentially well-equipped intruder in order to compromise a single resource-poor node. Following a successful attack, a compromised sensor node could then be used to launch such malicious activities as advertising false routing information, and launching DoS attacks from within the sensor network.

The combined threats introduced by increased physical security risk and severe resource constraints motivate the following design philosophy to achieve secure WSNs: assume that a well-equipped intruder can compromise individual sensor nodes, but secure the overall design of the WSN so that these intrusions can be tolerated and the network as a whole remains functioning despite such localized intrusions. More precisely, the objective is the design of an intrusion-tolerant WSN that has the property that a single compromised node can only disrupt a localized portion of the network, and cannot bring down the entire sensor network. This design objective of intrusion tolerance for secure WSNs must provide protection against two classes of attacks that could bring down an entire sensor network: DoS-type attacks and routing disruption attacks that propagate erroneous control packets containing false routing information throughout the network.

The focus of this chapter is on *routing security* in WSNs. Most of the currently existing routing protocols for WSNs make an optimization on the limited capabilities of the nodes and the application-specific nature of the network, but do not any the security aspects of the protocols. Although these protocols have not been designed with security as a goal, it is extremely important to analyze their security properties. When the defender has the liabilities of insecure wireless communication, limited node capabilities, and possible insider threats, and the adversaries can use powerful laptops with high energy and long range communication to attack the network, designing a secure routing protocol for WSNs is obviously a non-trivial task.

One aspect of sensor networks that complicates the design of a secure routing protocol is *in-network aggregation* (Shrivastava et al., 2004; Madden et al., 2002; Przydatck et al., 2003; Zhu et al., 2004a). In more conventional networks, a secure routing protocol is typically only required to guarantee message availability. Message integrity, authenticity, and confidentiality are handled at a higher layer by an end-to-end security mechanism such as SSH or SSL. End-to-end security is possible in more conventional networks because it is neither necessary nor desirable for intermediate routers to have access to the contents of messages. However, in sensor networks, in-network processing makes end-to-end security mechanism harder to deploy because intermediate nodes need direct access to the contents of the messages. Link layer security mechanisms can help mediate some of the resulting vulnerabilities, but it is not enough: we will now require much more from our protocols, and they must be designed with this in mind.

The organization of this chapter is as follows. In Section 2, we discuss the various resource constraints under which a typical WSN operates. In Section 3, various security requirements of such networks are identified. In section 4, a number of security vulnerabilities of WSNs are presented. Different types of attacks at various layers such as physical, link, network and transport layers are discussed in detail. In particular, various attacks at the network layers are described such as : (i) spoofed routing information (Karlof et al., 2003), (ii) selective packet forwarding (Karlof et al., 2003), (iii) sinkhole (Wood et al., 2002), (iv) Sybil (Newsome et al., 2004), (v) wormhole (Karlof et al., 2003), (vi) hello flood (Karlof et al., 2003), (vii) acknowledgment spoofing etc (Karlof et al., 2003). Section 5 presents a discussion on the defense mechanisms for DoS attacks at the network layer. In particular, schemes such as use of message authentication code (MAC) (Perrig et al., 2002), directional antenna-based defense (Hu et al., 2004a), packet leashes (Hu et al., 2004b), client puzzles (Aura et al., 2001) are discussed. Section 6 discusses secure broadcasting and multicasting techniques based on group key management protocols (Rafaeli et al., 2003) and directed diffusion-based mechanism (Di Pietro et al., 2003) etc. Section 7 presents some of the well-known existing secure routing protocols for WSNs such as *μTESLA* (Liu et al., 2004), INSENS (Deng et al., 2002b), SPINS (Perrig et al., 2002), TRANS (Tanachawiwat et al., 2003), and defense mechanisms against Sybil attack (Newsome et al., 2004; Chan, et al., 2003b; Eschenauer et al., 2002; Du et al., 2003), blackhole and grayhole (Sen et al., 2007b) attacks, a secure and energy-efficient routing protocol (Sen et al., 2010) are also discussed in detail. Finally, in conclusion, some future research directions are discussed.

In summary, the chapter makes the following contributions:
- It proposes threat models and security goals for secure routing in WSNs.
- It identifies various possible attacks on the network layer of a WSN sensor networks
- It demonstrates how attacks against ad-hoc wireless networks and peer-to-peer networks can be adapted into powerful attacks against WSNs.
- It presents a detailed security analysis of all the major routing protocols and energy conserving topology maintenance algorithms for WSNs.
- It presents various defense mechanisms to counter the well-known attacks on the routing protocols of WSNs.

## 2. Constraints in WSNs

A WSN consists of a large number of sensor nodes which are inherently resource-constrained. These nodes have limited processing capability, very low storage capacity, and constrained communication bandwidth. These limitations are due to limited energy and physical size of the sensor nodes. Due to these constraints, it is difficult to directly employ the conventional security mechanisms in WSNs. In order to optimize the conventional security algorithms for WSNs, it is necessary to be aware about the constraints of sensor nodes (Carman et al., 2000). The major constraints of a WSN are listed below.

(i) *Energy constraints*: Energy is the biggest constraint for a WSN. In general, energy consumption in sensor nodes can be categorized in three parts: (i) energy for the sensor transducer, (ii) energy for communication among sensor nodes, and (iii) energy for microprocessor computation. The study in (Hill et al., 2000) found that each bit transmitted in WSNs consumes about as much power as executing 800 to 1000 instructions. Thus, communication is more costly than computation in WSNs. Any message expansion caused by security mechanisms comes at a significant cost. Further, higher security levels in WSNs usually correspond to more energy consumption for cryptographic functions. Thus, WSNs could be divided into different security levels depending on energy cost (Slijepcevic et al., 2002; Yuan et al., 2002).

(ii) *Memory limitations:* A sensor is a tiny device with only a small amount of memory and storage space. Memory is a sensor node usually includes flash memory and RAM. Flash memory is used for storing downloaded application code and RAM is used for storing application programs, sensor data, and intermediate results of computations. There is usually not enough space to run complicated algorithms after loading the OS and application code. In the SmartDust project, for example, TinyOS consumes about 4K bytes of instructions, leaving only 4500 bytes for security and applications (Hill et al., 2000). A common sensor type- TelosB- has a 16-bit, 8 MHz RISC CPU with only 10K RAM, 48K program memory, and 1024K flash storage. The current security algorithms are therefore, infeasible in these sensors (Perrig et al., 2002).
(iii) *Unreliable communication:* Unreliable communication is another serious threat to sensor security. Normally the packet-based routing of sensor networks is based on connectionless protocols and thus inherently unreliable. Packets may get damaged due to channel errors or may get dropped at highly congested nodes. Furthermore, the unreliable wireless communication channel may also lead to damaged or corrupted packets. Higher error rate also mandates robust error handling schemes to be implemented leading to higher overhead. In certain situation even if the channel is reliable, the communication may not be so. This is due to the broadcast nature of wireless communication, as the packets may collide in transit and may need retransmission (Akyildiz et al., 2002).
(iv) *Higher latency in communication:* In a WSN, multi-hop routing, network congestion and processing in the intermediate nodes may lead to higher latency in packet transmission. This makes synchronization very difficult to achieve. The synchronization issues may sometimes be very critical in security as some security mechanisms may rely on critical event reports and cryptographic key distribution (Stankovic, 2003).
(v) *Unattended operation of networks:* In most cases, the nodes in a WSN are deployed in remote regions and are left unattended. The likelihood that a sensor encounters a physical attack in such an environment is therefore, very high. Remote management of a WSN makes it virtually impossible to detect physical tampering. This makes security in WSNs a particularly difficult task.

## 3. Security Requirements in WSNs

A WSN is a special type of network. It shares some commonalities with a typical computer network, but also exhibits many characteristics which are unique to it. The security services in a WSN should protect the information communicated over the network and the resources from attacks and misbehavior of nodes. The most important security requirements in WSN are listed below:
(i) *Data confidentiality*: The security mechanism should ensure that no message in the network is understood by anyone except the intended recipient. In a WSN, the issue of confidentiality should address the following requirements (Carman et al., 2000; Perrig et al., 2002): (i) a sensor node should not allow its readings to be accessed by its neighbors unless they are authorized to do so, (ii) key distribution mechanism should be extremely robust, (iii) public information such as sensor identities, and public keys of the nodes should also be encrypted in certain cases to protect against traffic analysis attacks.
(ii) *Data integrity*: The mechanism should ensure that no message can be altered by an entity as it traverses from the sender to the recipient.
(iii) *Availability*: This requirements ensures that the services of a WSN should be available always even in presence of an internal or external attacks such as a *denial of service* (DoS) attack. Different approaches have been proposed by researchers to achieve this goal. While some mechanisms make use of additional communication among nodes, others propose use of a central access control system to ensure successful delivery of every message to its recipient.
(iv) *Data freshness*: It implies that the data is recent and ensures that no adversary can replay old messages. This requirement is especially important when the WSN nodes use shared-keys for message communication, where a potential adversary can launch a replay attack using the old key as the new key is being refreshed and propagated to all the nodes in the WSN. A nonce or time-specific counter may be added to each packet to check the freshness of the packet.

(v) *Self-organization:* Each node in a WSN should be self-organizing and self-healing. This feature of a WSN also poses a great challenge to security. The dynamic nature of a WSN makes it sometimes impossible to deploy any pre-installed shared key mechanism among the nodes and the base station (Eschenauer et al., 2002). A number of key pre-distribution schemes have been proposed in the context of symmetric encryption (Chan et al., 2003b; Eschenauer et al., 2002; Hwang et al., 2004; Liu, et al., 2005a). However, for application of public-key cryptographic techniques an efficient mechanism for key-distribution is very much essential. It is desirable that the nodes in a WSN self-organize among themselves not only for multi-hop routing but also to carry out key management and developing trust relations.

(vi) *Secure localization*: In many situations, it becomes necessary to accurately and automatically locate each sensor node in a WSN. For example, a WSN designed to locate faults would require accurate locations of sensor nodes identifying the faults. A potential adversary can easily manipulate and provide false location information by reporting false signal strength, replaying messages etc., if the location information is not secured properly. The authors in (Capkun et al., 2006) have described a technique called *verifiable multi-lateration* (VM). In multi-lateration, the position of a device is accurately computed from a series of known reference points. The authors have used authenticated ranging and distance bounding to ensure accurate location of a node. Because of the use of distance bounding, an attacking node can only increase its claimed distance from a reference point. However, to ensure location consistency, the attacker would also have to prove that its distance from another reference point is shorter. As it is not possible for the attacker to prove this, it is possible to detect the attacker. In (Lazos et al., 2005), the authors have described a scheme called *secure range-independent localization* (SeRLoC). The scheme is a decentralized range-independent localization scheme. It is assumed that the locators are trusted and cannot be compromised by any attacker. A sensor computes its location by listening to the beacon information sent by each locator which includes the locator's location information. The beacon messages are encrypted using a shared global symmetric key that is pre-distributed in the sensor nodes. Using the information from all the beacons that a sensor node receives, it computes its approximate location based on the coordinates of the locators. The sensor node then computes an overlapping antenna region using a majority vote scheme. The final location of the sensor node is determined by computing the center of gravity of the overlapping antenna region.

(vii) *Time synchronization*: Most of the applications in sensor networks require time synchronization. Any security mechanism for WSN should also be time-synchronized. A collaborative WSN may require synchronization among a group of sensors. In (Ganeriwal et al., 2005), the authors have proposed a set of secure synchronization protocols for multi-hop sender-receiver and group synchronization.

(viii) *Authentication*: It ensures that the communicating node is the one that it claims to be. An adversary can not only modify data packets but also can change a packet stream by injecting fabricated packets. It is, therefore, essential for a receiver to have a mechanism to verify that the received packets have indeed come from the actual sender node. In case of communication between two nodes, data authentication can be achieved through a *message* *authentication code* (MAC) computed from the shared secret key among the nodes. A number of authentication schemes for WSNs have been proposed by researchers. Most of these schemes are for secure routing and reliable packet. Some of these schemes will be discussed in Section 5.

## 4. Security Vulnerabilities in WSNs

Wireless Sensor Networks are vulnerable to various types of attacks. These attacks are mainly of three types (Shi et al., 2004):
(i) *Attacks on network availability*: attacks on availability of WSN are often referred to as DoS attacks.
(ii) *Attacks on secrecy and authentication*: standard cryptographic techniques can protect the secrecy and authenticity of communication channels from outsider attacks such as eavesdropping, packet replay attacks, and modification or spoofing of packets.
(iii) *Stealthy attack against service integrity*: in a stealthy attack, the goal of the attacker is to make the network accept a false data value. For example, an attacker compromises a sensor node and injects a false data value through that sensor node.
In these attacks, keeping the sensor network available for its intended use is essential. DoS attacks against WSNs may permit real-world damage to the health and safety of people (Wood et al., 2002). The DoS attack usually refers to an adversary's attempt to disrupt, subvert, or destroy a network. However, a DoS attack can be any event that diminishes or eliminates a network's capacity to perform its expected functions (Wood et al., 2002).

### 4.1 Denial of Service Attacks

Wood and Stankovic have defined a DoS attack as an event that diminishes or attempts to reduce a network's capacity to perform its expected function (Wood et al., 2002). There are several standard techniques existing in the literature to cope with some of the more common denial of service attacks, although in a broader sense, development of a generic defense mechanism against DoS attacks is still an open problem. Moreover, most of the defense mechanisms require high computational overhead and hence not suitable for resource-constrained WSNs. Since DoS attacks in WSNs can sometimes prove very costly, researchers have spent a great deal of effort in identifying various types of such attacks, and devising strategies to defend against them. Some of the important types of DoS attacks at different layers of WSNs are discussed below:
**(a) Physical layer attacks**: The physical layer is responsible for frequency selection, carrier frequency generation, signal detection, modulation, and data encryption (Akyildiz et al. 2002). As with any radio-based medium, the possibility of jamming is there. The nodes in WSNs may be deployed in hostile or insecure environments, where an attacker has the physical access. Two types of attacks in physical layer are (i) jamming and (ii) tampering.
(i) *Jamming:* it is a type of attack which interferes with the radio frequencies that the nodes use in a WSN for communication (Wood et al., 2002; Shi et al., 2004). A jamming source may be powerful enough to disrupt the entire network. Even with less powerful jamming sources, an adversary can potentially disrupt communication in the entire network by strategically distributing the jamming sources. Even an intermittent jamming may prove detrimental as the message communication in a WSN may be extremely time-sensitive (Wood et al., 2002).

(ii) *Tampering*: sensor networks typically operate in outdoor environments. Due to unattended and distributed nature, the nodes in a WSN are highly susceptible to physical attacks (Wang et al., 2004a). The physical attacks may cause irreversible damage to the nodes. The adversary can extract cryptographic keys from the captured node, tamper with its circuitry, modify the program codes, or even replace it with a malicious sensor (Wang et al., 2005). It has been shown that sensor nodes such as MICA2 motes can be compromised in less than one minute time (Hartung, et al., 2004).

**(b) Link layer attacks**: The link layer is responsible for multiplexing of data-streams, data frame detection, medium access control, and error control (Akyildiz et al., 2002). Attacks at this layer include purposefully created collisions, resource exhaustion, and unfairness in allocation.

A collision occurs when two nodes attempt to transmit on the same frequency simultaneously (Wood et al., 2002). When packets collide, they are discarded and need to re-transmitted. An adversary may strategically cause collisions in specific packets such as ACK control messages. A possible result of such collisions is the costly exponential back-off. The adversary may simply violate the communication protocol, and continuously transmit messages in an attempt to generate collisions. Repeated collisions can also be used by an attacker to cause resource exhaustion (Wood et al., 2002). For example, a naïve link layer implementation may continuously attempt to retransmit the corrupted packets. Unless these retransmissions are detected early, the energy levels of the nodes would be exhausted quickly. Unfairness is a weak form of DoS attack (Wood et al., 2002). An attacker may cause unfairness by intermittently using the above link layer attacks. In this case, the adversary causes degradation of real-time applications running on other nodes by intermittently disrupting their frame transmissions.

**(c) Network layer attacks:** The network layer of WSNs is vulnerable to the different types of attacks such as: spoofed routing information, selective packet forwarding, sinkhole, Sybil, wormhole, blackhole, hello flood, Byzantine attack, information disclosure, resource depletion attack, acknowledgment spoofing, routing table overflow, route poisoning, rushing attack etc. These attacks are described briefly in the following:

(i) *Spoofed routing information*: the most direct attack against a routing protocol is to target the routing information in the network. An attacker may spoof, alter, or replay routing information to disrupt traffic in the network (Karlof et al., 2003). These disruptions include creation of routing loops, attracting or repelling network traffic from selected nodes, extending or shortening source routes, generating fake error messages, causing network partitioning, and increasing end-to-end latency.

(ii) *Selective forwarding*: in a multi-hop network like a WSN, for message communication all the nodes need to forward messages accurately. An attacker may compromise a node in such a way that it selectively forwards some messages and drops others (Karlof et al., 2003).

(iii) *Sinkhole*: In a sinkhole attack, an attacker makes a compromised node look more attractive to its neighbors by forging the routing information (Karlof et al., 2003; Wood et al., 2002; Newsome et al., 2004). The result is that the neighbor nodes choose the compromised node as the next-hop node to route their data through. This type of attack makes selective forwarding very simple as all traffic from a large area in the network would flow through the compromised node.

(iv) *Sybil attack*: it is an attack where one node presents more that one identity in a network. It was originally described as an attack intended to defeat the objective of redundancy mechanisms in distributed data storage systems in peer-to-peer networks (Douceur, 2002). Newsome et al. describe this attack from the perspective of a WSN (Newsome et al., 2004). In addition to defeating distributed data storage systems, the Sybil attack is also effective against routing algorithms, data aggregation, voting, fair resource allocation, and foiling misbehavior detection. Regardless of the target (voting, routing, aggregation), the Sybil algorithm functions similarly. All of the techniques involve utilizing multiple identities. For instance, in a sensor network voting scheme, the Sybil attack might utilize multiple identities to generate additional "votes". Similarly, to attack the routing protocol, the Sybil attack would rely on a malicious node taking on the identity of multiple nodes, and thus routing multiple paths through a single malicious node.

(v) *Wormhole*: a wormhole is low latency link between two portions of a network over which an attacker replays network messages (Karlof et al., 2003). The attacker receives packets at one location in the network, and tunnels them to another location in the network, where the packets are resent into the network. The tunnel between the two colluding attackers is known as the *wormhole*. This link may be established either by a single node forwarding messages between two adjacent but otherwise non-neighboring nodes or by a pair of nodes in different parts of the network communicating with each other. The latter case is closely related to sinkhole attack as an attacking node near the base station can provide a one-hop link to that base station via the other attacking node in a distant part of the network. Due to the broadcast nature of the radio channel, the attacker can create a wormhole link even for packets which are not addressed to it. If proper security mechanisms are not deployed to defend against such attacks, routing in WSN may be impossible.

(vi) *Blackhole* and *Grayhole*: in this attack, a malicious node falsely advertises good paths (e.g. the shortest path or the most stable path) to the destination node during the path-finding process (in reactive routing protocols), or in the route updates messages (in proactive routing protocols). The intention of the malicious node could be to hinder the path-finding process or to intercept all data packets being sent to the destination node concerned. A more delicate form of this attack is known as the grayhole attack, where the malicious node intermittently drops the data packets thereby making its detection even more difficult.

(vii) *Hello flood*: most of the protocols that use Hello packets make the naïve assumption that receiving such a packet implies that the sender is within the radio range of the receiver. An attacker may use a high-powered transmitter to fool a large number of nodes and make them believe that they are within its neighborhood (Karlof et al., 2003). Subsequently, the attacker node falsely broadcasts a shorter route to the base station, and all the nodes which received the Hello packets, attempt to transmit to the attacker node. However, these nodes are out of the radio range of the attacker.

(viii)*Byzantine attack*: in this attack, a compromised node or a set of compromised nodes works in collusion and carries out attacks such as creating routing loops, forwarding packets in non-optimal routes, and selectively dropping packets (Awerbuch et al., 2002). Byzantine attacks are very difficult to detect, since under such attacks the networks usually do not exhibit any abnormal behavior.

(ix) *Information disclosure*: a compromised node may leak confidential or important information to unauthorized nodes in the network. Such information may include information regarding the network topology, geographic location of nodes, or optimal routes to authorized nodes in the network.

(x) *Resource depletion attack*: in this type of attack, a malicious node tries to deplete resources of other nodes in the network. The typical resources that are targeted are: battery power, bandwidth, and computational power. The attacks could be in the form of unnecessary requests for routes, very frequent generation of beacon packets, or forwarding of stale packets to other nodes.

*Acknowledgment spoofing*: some routing algorithms for WSNs require transmission of acknowledgment packets. An attacking node may overhear packet transmissions from its neighboring nodes and spoof the acknowledgments thereby providing false information to the nodes (Karlof et al., 2003). In this way, the attacker is able to disseminate wrong information about the status of the nodes.

(xi) *Attacks on routing protocols*: most of the routing protocols for WSNs are vulnerable to various types of attacks. Some of these attacks are listed below.

- *Routing table overflow*: in this type of attack, an adversary node advertises routes to non-existent nodes, to the authorized node present in the network. The main objective of such an attack is to cause an overflow of the routing tables, which would in turn prevent the creation of entries corresponding to new routes to authorized nodes. Proactive routing protocols are more vulnerable to this attack compared to reactive routing protocols.
- *Routing table poisoning*: in this case, the compromised nodes in the network send fictitious routing updates or modify genuine route update packets sent to other honest nodes. Routing table poisoning may result in sub-optimal routing, congestion in some portions of the network, or even make some parts of the network inaccessible.
- *Packet replication*: in this attack, an adversary node replicates stale packets. This consumes additional bandwidth and battery power and other resources available to the nodes and also causes unnecessary confusion in the routing process.
- *Route cache poisoning*: in reactive (i.e. on-demand) routing protocols such as ad hoc on-demand distance vector (AODV) (Perkins, et al., 1999), each node maintains a route cache which holds information regarding routes that have become known to the node in the recent past. Similar to routing table poisoning, an adversary can also poison the route cache to achieve similar objectives.
- *Rushing attack*: on-demand routing protocols that use *duplicate suppression* during the route discovery process are vulnerable to this attack (Hu et al., 2003b). An adversary node which receives a *routerequest* packet from the source node floods the packet quickly throughout the network before other nodes which also receive the same *routerequest* packet can react. Nodes that receive the legitimate *routerequest* packets assume those packets to be duplicates of the packet already received through the adversary node and hence discard those packets. Any route discovered by the source node would contain the adversary node as one of the intermediate nodes. Hence, the source node would not be able to find secure routes, that is, routes that do not include the adversary node. It is extremely difficult to detect such attacks in WSNs.

**(d) Transport layer attacks:** The attacks that can be launched on the transport layer in a WSN are flooding attack and de-synchronization attack.

(i) *Flooding*: Whenever a protocol is required to maintain state at either end of a connection, it becomes vulnerable to memory exhaustion through flooding (Wood et al., 2002). An attacker may repeatedly make new connection request until the resources required by each connection are exhausted or reach a maximum limit. In either case, further legitimate requests will be ignored.

(ii) *De-synchronization*: De-synchronization refers to the disruption of an existing connection (Wood et al., 2002). An attacker may, for example, repeatedly spoof messages to an end host causing the host to request the retransmission of missed frames. If timed correctly, an attacker may degrade or even prevent the ability of the end hosts to successfully exchange data causing them instead to waste energy attempting to recover from errors which never really exist. The possible DoS attacks and the corresponding countermeasures are listed in Table 1.

| Layer | Attacks | Defense |
|---|---|---|
| Physical | Jamming | Spread-spectrum, priority messages, lower duty cycle, region mapping, mode change |
| Link | Collision | Error-correction code |
|  | Exhaustion | Rate limitation |
|  | Unfairness | Small frames |
| Network | Spoofed routing information & selective forwarding | Egress filtering, authentication, monitoring |
|  | Sinkhole | Redundancy checking |
|  | Sybil | Authentication, monitoring, redundancy |
|  | Wormhole | Authentication, probing |
|  | Hello Flood | Authentication, packet leashes by using geographic and temporal info |
|  | Ack. flooding | Authentication, bi-directional link authentication verification |
| Transport | Flooding | Client puzzles |
|  | De-synchronization | Authentication |

Table 1. Various attacks on WSNs and their countermeasures (Wang et al., 2006)

### 4.2 Attacks on Secrecy and Authentication
There are different types of attacks under this category as discussed below.

*(i) Node replication attack*: In a node replication attack, an attacker attempts to add a node to an existing WSN by replicating (i.e. copying) the node identifier of an already existing node in the network (Parno et al., 2005). A node replicated and joined in the network in this manner can potentially cause severe disruption in message communication in the WSN by corrupting and forwarding the packets in wrong routes. This may also lead to network partitioning, communication of false sensor readings etc. In addition, if the attacker gains physical access to the entire network, it is possible for him to copy the cryptographic keys and use these keys for message communication from the replicated node. The attacker can also place the replicated node in strategic locations in the network so that he could easily manipulate a specific segment of the network, possibly causing a network partitioning.

*(ii) Attacks on privacy*: Since WSNs are capable of automatic data collection through efficient and strategic deployment of sensors, these networks are also vulnerable to potential abuse of these vast data sources. Privacy preservation of sensitive data in a WSN is particularly difficult challenge (Gruteser et al., 2003). Moreover, an adversary may gather seemingly innocuous data to derive sensitive information if he knows how to aggregate data collected from multiple sensor nodes. This is analogous to the *panda hunter* problem, where the hunter can accurately estimate the location of the panda by monitoring the traffic (Ozturk et al., 2004).

The privacy preservation in WSNs is even more challenging since these networks make large volumes of information easily available through remote access mechanisms. Since the adversary need not be physically present to carryout the surveillance, the information gathering process can be done anonymously with a very low risk. In addition, remote access allows a single adversary to monitor multiple sites simultaneously (Chan et al., 2003a). Following are some of the common attacks on sensor data privacy (Gruteser et al., 2003, Chan et al., 2003a):

(iii) *Eavesdropping and passive monitoring*: This is the most common and the easiest form of attack on data privacy. If the messages are not protected by cryptographic mechanisms, the adversary could easily understand the contents. Packets containing control information in a WSN convey more information than accessible through the location server, Eavesdropping on these messages prove more effective for an adversary.

(iv) *Traffic analysis*: In order to make an effective attack on privacy, eavesdropping should be combined with a traffic analysis. Through an effective analysis of traffic, an adversary can identify some sensor nodes with special roles and activities in a WSN. For example, a sudden increase in message communication between certain nodes signifies that those nodes have some specific activities and events to monitor. Deng et al. have demonstrated two types of attacks that can identify the base station in a WSN without even underrating the contents of the packets being analyzed in traffic analysis (Deng et al., 2004).

(v) *Camouflage*: An adversary may compromise a sensor node in a WSN and later on use that node to masquerade a normal node in the network. This camouflaged node then may advertise false routing information and attract packets from other nodes for further forwarding. After the packets start arriving at the compromised node, it starts forwarding them to strategic nodes where privacy analysis on the packets may be carried out systematically.

It may be noted from the above discussion that WSNs are vulnerable to a number of attacks at all layers of the TCP/IP protocol stack. However, as pointed out by authors in (Perrig et al., 2004), there may be other types of attacks possible which are not yet identified. Securing a WSN against all these attacks may be a quite challenging task.

## 5. Network Layer Defense on DoS Attacks

A countermeasure against spoofing and alteration is to append a *message authentication code* (MAC) after the message. By adding a MAC to the message, the receivers can verify whether the messages have been spoofed or altered. To defend against replayed information, counters or time-stamps may be introduced in the messages (Perrig et al., 2002). A possible defense against selective forwarding attack is using multiple paths to send data (Karlof et al., 2003). A second defense is to detect the malicious node or assume it has failed and seek an alternative route.

Hu et al. have proposed a novel and generic mechanism called *packet leashes* for detecting and defending against wormhole attacks (Hu et al., 2004b). As mentioned in Section 4.1, in a wormhole attack, a malicious node eavesdrops on a series of packets, then tunnels them through a path in the network, and replays them. This is done in order to make a false representation of the distance between the two colluding nodes. It is also used, more generally, to disrupt the routing protocol by misleading the neighbor discovery process (Karlof et al., 2003). Hu et al. have presented a mechanism that employs directional antenna to combat wormhole attack (Hu et al., 2004a). Wang and Bhargava have used a visualization approach to detect wormholes in a WSN (Wang et al., 2004b). In the mechanism proposed by the authors, a distance estimation is made between all the sensor nodes in a neighborhood. Using multi-dimensional scaling, a virtual layout of the network is then computed, and a surface smoothing strategy is used to adjust the round-off errors. Finally, the shape of the resulting virtual network is analyzed. If any wormhole exists, the shape of the network will bend and curve towards the wormhole, otherwise the network will appear flat.

To defend against flooding DoS attack at the transport layer, Aura et al. have proposed a mechanism using client puzzles (Aura et al., 2001). The main idea is that each connecting client should demonstrate its commitment to the connection by solving a puzzle. As an attacker in most likelihood, does not have infinite resource, it will be impossible for him to create new connections fast enough to cause resource starvation on the serving node.

A possible defense against de-synchronization attack on the transport layer is to enforce a mandatory requirement of authentication of all packets communicated between nodes (Wood et al., 2002). If the authentication mechanism is secure, an attacker will be unable to send any spoofed messages to any destination node.

Some mechanisms for secure multicasting and broadcasting in WSNs are discussed in the following sub-section.

## 6. Secure Broadcasting and Multicasting Protocols for WSNs

Multicasting and broadcasting techniques are used primarily to reduce the communication and management overhead of sending a single message to multiple receivers. In order to ensure that only legitimate group members receive the multicast and broadcast communication, appropriate authentication and encryption mechanisms must be in place. To handle this problem, several key management schemes have been devised: centralized group key management protocols, decentralized key management protocols, and distributed key management protocols (Rafaeli et al., 2003). First, we will discuss some generic security mechanisms for multicast and broadcast communication in wireless networks. Then we will present some of the well-known propositions specific to WSNs.

In the case of the centralized group key management protocols, a central authority is used to maintain the group. Decentralized management protocols, however, divide the task of group management amongst multiple nodes. In distributed key management protocols, the key management activity is distributed among a set of nodes rather than on a single node. In some cases, the entire group of nodes is responsible for key management (Rafaeli et al., 2003).

An efficient way to distribute keys in a network is to use a logical key tree. Such techniques essentially fall under the category of centralized key management protocols. Some schemes have been developed for WSNs based on logical key tree technique (Di Pietro et al., 2003; Lazos et al., 2002; Lazos et al., 2003). While centralized solutions are not always the most efficient ones, these mechanisms may sometimes be very effective for WSNs, as relatively heavier computations can be usually carried out in powerful base stations.

Di Pietro et al. have proposed a directed diffusion-based multicast mechanism for WSNs that utilizes a logical key hierarchy (Di Pietro et al., 2003). In the logical hierarchy, a central key distributor is at the root of a tree, and the nodes in the network are the leaf level. The internal nodes of tree contain keys that are used in the re-keying process. The directed diffusion is an energy-efficient data dissemination technique for WSNs (Intanagonwiwat et al., 2000). In directed diffusion, a query is transformed into an interest and then diffused throughout the network. The source node then starts collecting data from the network based on the propagated interest. The dissemination technique also sets up certain gradients designed to draw events toward the interest. The collected data is then sent back to the source along the reverse path of the interest propagation. The directed diffusion-based logical key hierarchy scheme as proposed by Di Pietro et al. allows nodes to join and leave groups. The key hierarchy is used to effectively re-establish keys for the nodes below the node that has left the group. When a node declares its intension to join a group, a key set is generated for the new node based on the keys within the existing key hierarchy.

Kaya et al. discuss the problem of multicast group management in (Kaya et al., 2003). In their proposition, the nodes in a network are grouped based on their locality and a security tree is constructed on the groups.

Lazos and Poovendran have presented a tree-based key distribution scheme that is similar to the directed diffusion-based logical key hierarchy proposed by Di Pietro et al. (Lazos et al., 2003). In their proposed scheme, a routing-aware tree is constructed in which the leaf nodes are assigned keys based on all relay nodes above them. As the scheme takes advantage of routing information for construction the key hierarchy, it is more energy-efficient than routing schemes that arbitrarily arrange nodes into a routing tree. The authors have also proposed a greedy routing-aware key distribution algorithm.

In (Lazos et al., 2003), the authors have proposed a mechanism that uses geographic location information (e.g. GPS data) for construction of a logical key hierarchy for secure multicast communication. The nodes, based on the geographical location information, are grouped into different clusters. The nodes within a cluster are able to reach each other with a single hop communication. Using the cluster information, a key hierarchy is constructed in a manner similar to that proposed in (Lazos et al., 2002).

## 7. Secure Routing Protocols for WSNs

Many routing protocols have been proposed for WSNs. These protocols can be divided into three broad categories according to the network structure: (i) flat-based routing, (ii) hierarchical-based routing, and (iii) location-based routing (Al-Karaki et al., 2004). In flat-based routing, all nodes are typically assigned equal roles or functionality. In hierarchical-based routing, nodes play different roles in the network. In location-based routing, sensor node positions are used to route data in the network. One common location-based routing protocol is GPSR (Karp et al., 2000). It allows nodes to send packets to a region rather than to a particular node. All these routing protocols are vulnerable to various types of attacks such as selective forwarding, sinkhole attack etc as mentioned in Section 4. An elaborate discussion on various types of attacks on the routing protocols in WSNs is given in (Karlof et al., 2003).

The goal of a secure routing protocol for a WSN is to ensure the integrity, authentication, and availability of messages. Most of the existing secure routing algorithms for WSNs are all based on symmetric key cryptography except the work in (Du et al., 2005), which is based on public key cryptography. In the following sub-sections, some of the existing secure routing protocols for WSNs are discussed in detail.

**7.1 Micro TESLA Protocol**

The "micro" version of the *Timed, Efficient, Streaming, Loss-tolerant Authentication* (μTESLA) protocol (Perrig et al., 2002) and its extensions (Liu et al., 2003; Liu et al. 2004) have been proposed to provide broadcast authentication for sensor networks. μTESLA is broadcast authentication mechanism which was proposed by Perrig et al. for the SPINS protocol (Perrig et al., 2002). μTESLA introduces asymmetry through a delayed disclosure of symmetric keys resulting in an efficient broadcast authentication scheme. For its operation, it requires the base station and the sensor nodes to be loosely synchronized. In addition, each node must know an upper bound on the maximum synchronization error.

To send an authenticated packet, the base station simply computes a MAC on the packet with a key that is secret at that point of time. When a node gets a packet, it can verify that the corresponding MAC key was not yet disclosed by the base station. Because a receiving node is assured that the MAC key is known only to the base station, the receiving node is assured that no adversary could have altered the packet in transit. The node stores the packet in a buffer. At the time of key disclosure, the base station broadcasts the verification key to all its receivers. When a node receives the disclosed key, it can easily verify the correctness of the key. If the key is correct, the node can now use it to authenticate the packet stored in its buffer.

Each MAC is a key from the key chain, generated by a public one-way function $F$. To generate the one-way key chain, the sender chooses the last key $K_n$ from the chain, and repeatedly applies $F$ to compute all other keys: $K_i = F(K_{i+1})$.

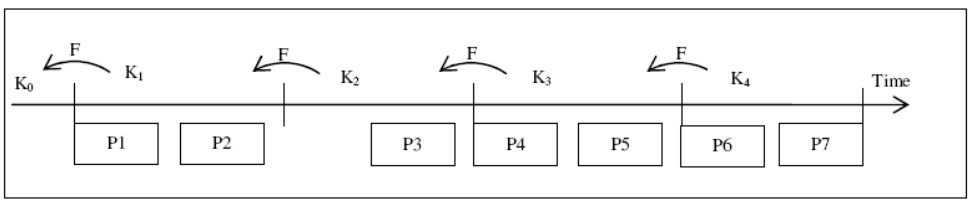

Fig. 1. Time-released key chain for source authentication (Wang et al. 2006)

Fig. 1 shows an example of μTESLA. The receiver node is loosely time synchronized and knows $K_0$ in an authenticated way. Packets $P_1$ and $P_2$ sent in interval 1 contain a MAC with a key $K_1$. Packet $P_3$ has a MAC using key $K_2$. If $P_4$, $P_5$, and $P_6$ are all lost, as well as the packet that disclosed the key $K_1$, the receiver cannot authenticate $P_1$, $P_2$, and $P_3$. In interval 4, the base station broadcasts the key $K_2$, which the nodes authenticate by verifying $K_0 = F(F(K_2))$, and hence know also $K_1 = F(K_2)$, so they can authenticate packets $P_1$, $P_2$ with $K_1$, and $P_3$ with $K_2$. SPINS limits the broadcasting capability to only the base station. If a node wants to broadcast authenticated data, the node has to broadcast the data through the base station. The data is first sent to the base station in an authenticated way. It is then broadcasted by the base station.

To bootstrap a new receiver, μTESLA depends on a point-to-point authentication mechanism in which a receiver sends a request message to the base station and the base station replies with a message containing all the necessary parameters. It may be noted that μTESLA requires the base station to unicast initial parameters to individual sensor nodes, and thus incurs a long delay to boot up a large-scale sensor network. Liu and Ning have proposed a multi-level key chain scheme for broadcast authentication to overcome this deficiency (Liu et al., 2003; Liu et al. 2004).

The basic idea in (Liu et al., 2003; Liu et al., 2004) is to predetermine and broadcast the initial parameters required by μTESLA instead of using unicast-based message transmission. The simplest way is to pre-distribute the μTESLA parameters with a master key during the initialization of the sensor nodes. As a result, all sensor nodes have the key chain commitments and other necessary parameters once they are initialized, and are ready to use μTESLA as long as the staring time has passed. Furthermore, the authors have introduced a multi-level key chain scheme, in which the higher key chains are used to authenticate the commitments of the lower-level ones. However, the multi-level key chain suffers from possible DoS attacks during commitment distribution stage. Further, none of the μTESLA or multi-level key chain schemes is scalable in terms of the number of senders. In (Liu et al., 2005b), a practical broadcast authentication protocol has been proposed to support a potentially large number of broadcast senders using μTESLA as a building block.

μTESLA provides broadcast authentication for base stations, but is not suitable for local broadcast authentication. This is because μTESLA does not provide immediate authentication. For every received packet, a node has to wait for one μTESLA interval to receive the MAC key used in computing the MAC for the packet. As a result, if μTESLA is used for local broadcast authentication, a message traversing $l$ hops will take at least $l$ μTESLA intervals to arrive at the destination. In addition, a sensor node has to buffer all unverified packets. Both the latency and the storage requirements limit the scheme for authenticating infrequent messages broadcast by the base station. Zhu et al. have proposed a one-way key chain scheme for one-hop broadcast authentication (Zhu et al., 2004b). The mechanism is known as LEAP. In this scheme, every node generates a one-way key chain of certain length and then transmits the commitment (i.e., first key) of the key chain to each neighbor, encrypted with their pair-wise shared key. Whenever a node has a message to send, it attaches to the message to the next authenticated key in the key chain. The authenticated keys are disclosed in reverse order to their generation. A receiving neighbor can verify the message based on the commitment or an authenticated key it received from the sending node more recently.

**7.2 Intrusion Tolerant Routing Protocol in WSNs**

Deng et al. have proposed an *intrusion tolerant routing protocol in wireless sensor networks* (INENS) that adopts a routing-based approach to security in WSNs (Deng et al., 2002b). It constructs routing tables in each node, bypassing malicious nodes in the network. The protocol can not totally rule out attack on nodes, but it minimizes the damage caused to the network. The computation, communication, storage, and bandwidth requirements at the nodes are reduced, but at the cost of greater computation and communication at the base station. To prevent DoS attacks, individual nodes are not allowed to broadcast to the entire network. Only the base station is allowed to broadcast, and the base station is authenticated using one-way hash function so as to prevent a possible masquerading by a malicious node. Control information pertaining to routing is authenticated by the base station in order to prevent injection of false routing data. The base station computes and disseminates routing tables, since it does not have computational and energy constraints. Even if an intruder takes over a node and does not forward packets, INSENS uses redundant multi-path routing, so that the destination can still reach without passing through the malicious node.

INSENS has two phases: *route discovery* and *data forwarding*. During the route discovery phase, the base station sends a request message to all nodes in the network by multi-hop forwarding. Any node receiving a request message records the identity of the sender and sends the message to all its immediate neighbors if it has not already done so. Subsequent request messages are used to identify the senders as neighbors, but repeated flooding is not performed. The nodes respond with their local topology by sending feedback messages. The integrity of the messages is protected using encryption by a shared key mechanism. A malicious node can inflict damage only by not forwarding packets, but the messages are sent through different neighbors, so it is likely that it reaches a node by at least one path. Hence, the effect of malicious nodes is not totally eliminated, but it is restricted to only a few downstream nodes in the worst case. Malicious nodes may also send spurious messages and cause battery drain for a few downstream nodes. Finally, the base station calculates forwarding tables for all nodes, with two independent paths for each node, and sends them to the nodes. The second phase of data forwarding takes place based on the forwarding tables computed by the base station.

## 7.3 Security Protocols for Sensor Networks

SPINS is a suite of security protocols optimized for sensor networks (Perrig et al., 2002). SPINS includes two building blocks: (i) *secure network encryption protocol* (SNEP) and (ii) μTESLA protocol. SNEP provides data confidentiality, two-party data authentication, and data freshness for peer-to-peer communication (node to base station). *μTESLA* provides authenticated broadcast as discussed already.

SPINS assumes that each node is pre-distributed with a master key $K$ which is shared with the base station at its time of creation. All the other keys, including a key $K_{encr}$ for encryption, a key $K_{mac}$ for MAC generation, and a key $K_{rand}$ for random number generation are derived from the master key using a string one-way function. SPINS uses RC5 protocol for confidentiality. If $A$ wants to send a message to base station $B$, the complete message $A$ sends to $B$ is:

$$A \rightarrow B : D_{<K_{encr}C>}, \text{MAC}(K_{mac}, C \mid D)_{<K_{encr}C>}$$

In the above expression, $D$ is the transmitted data and $C$ is a shared counter between the sender and the receiver for the block cipher in counter mode. The counter $C$ is incremented after each message is sent and received in both the sender and the receiver side. SNEP also provides a counter exchange protocol to synchronize the counter value in both sides.

SNEP provides the flowing properties:

(i) *Semantic security*: the counter value is incremented after each message and thus the same message is encrypted differently each time.
(ii) *Data authentication*: a receiver can be assured that the message originated from the claimed sender if the MAC verification produces positive results.
(iii) *Replay protection*: the counter value in the MAC prevents replaying old messages by an adversary.
(iv) *Weak freshness*: SPINS identifies two types of freshness. Weak freshness provides partial message ordering and carries no delay information. Strong freshness provides a total order on a request-response pair and allows delay estimation. IN SNEP, the counter maintains a message ordering in the receiver side and yields weak freshness. SNEP guarantees weak freshness only, since there is no guarantee to node *A* that a message was created by node *B* in response to an event in node *A*.
(v) *Low communication overhead*: the counter state is kept at each endpoint and need not be sent in each message.

### 7.4 A Secure Protocol for Defending Cooperative Grayhole Attack

As mentioned in Section 4.1, blackhole and grayhole are two attacks that can severely disrupt routing in WSNs. A blackhole attack typically has two phases. In the first phase, the malicious node exploits the ad hoc routing protocol such as AODV (Perkins et al., 1999) to advertise itself as having a valid route to a destination node, with the intention of intercepting packets, even though the route is spurious. In the second phase, the attacker node drops the intercepted packets without forwarding them.

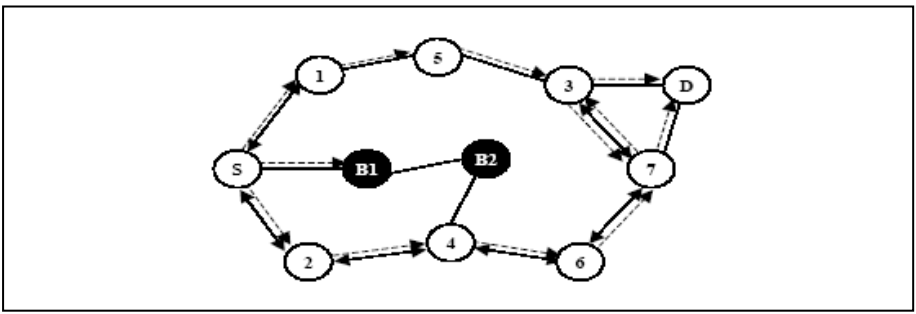

Fig. 2. Network flooding by RREQ and propagation of RREP (Deng et al., 2002a)

In the standard AODV protocol, when the source node *S* (Fig. 2) wants to communicate with the destination node *D*, the source node *S* broadcasts the *route request* (RREQ) packet. Each neighboring active node updates its routing table with an entry for the source node *S*, and checks if it is the destination node or whether it has the current route to the destination node. If an intermediate node does not have the current route to the destination node, it updates the RREQ packet by increasing the hop count and floods the network with the RREQ to the destination node *D* until it reaches node *D* or any other intermediate node that has the current route to *D*. The destination node *D* or any intermediate node that has the current route to *D*, initiates a *route reply* (RREP) in the reverse direction. Node *S* starts sending data packets to the neighboring node that responded first, and discards the other responses. This works fine when the network has no malicious nodes.

In (Deng et al., 2002a), authors have proposed a solution to identify and isolate a single blackhole node. However, the security threat arising out of the situation where multiple blackhole nodes act in coordination has not been addressed. For example, in Fig. 2, when more than one blackhole nodes are acting in coordination with each other, the first black hole node $B_1$ refers to one of its partners $B_2$ as the next hop. In the mechanism proposed in (Deng et al., 2002a), the source node $S$ sends *further request* (FRq) to $B_2$ through a different route ($S \rightarrow 2 \rightarrow 4 \rightarrow B_2$) other than via $B_1$. Node $S$ asks $B_2$ if it has a route to node $B_1$ and a route to destination node $D$. Since $B_2$ is cooperating with $B_1$, its *further reply* (FRp) will be '*yes*' to both the queries. Node $S$ starts sending the data packets assuming that the route $S \rightarrow B_1 \rightarrow B_2$ is secure. However, in reality, the packets are intercepted and then dropped by the node $B_1$ and the security of the network is compromised.

Sen et al. have proposed a security mechanism that can detect cooperative grayhole attacks in a wireless ad hoc and sensor network (Sen et al., 2007b). As mentioned in Section 4.1, detection of grayholes is more difficult than detection of blackholes since these nodes drop packets intermittently and change their behavior frequently so as to avoid detection. In the proposed mechanism, each node in the network collects the data forwarding information in its neighborhood and stores it in a table known as the *data routing information* (DRI) table.

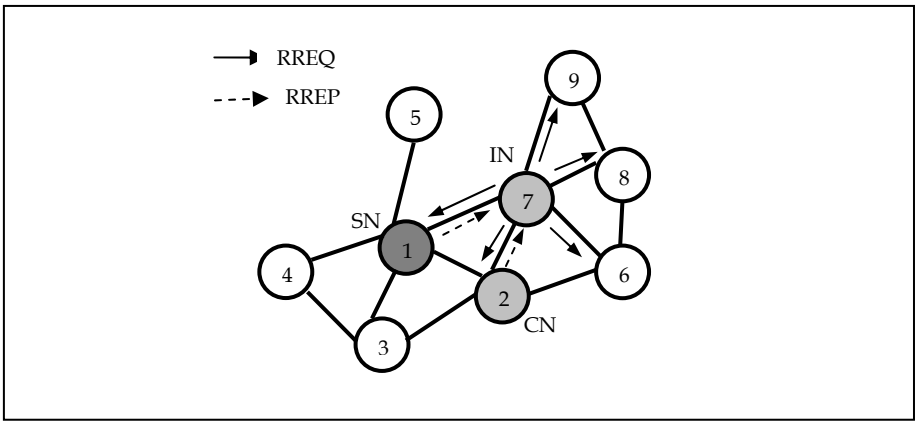

Fig. 3. The topology of a wireless ad hoc and sensor network (Sen et al., 2007b)

The DRI table of node 7 in Fig. 3 is shown in Table 2. In its DRI table node 7 maintains packet routing information of its neighbor nodes 1, 2, 6, 8, and 9. An entry '1' for a node under the column '*From*' implies that node 7 has forwarded data packet coming from that node and an entry '1' for a node under the column '*Through*' implies that node 7 has forwarded data packets to that node. Thus, as per Table 2, node 7 has neither forwarded any data packet from node 1 nor it has forwarded any data packet to node 1. However, node 7 has forwarded data packets to node 2 and also has forwarded data packets that have come from node 2. In this way, each node constructs its DRI table and maintains it. After a certain threshold time interval, each node identifies its neighbors with which it has not interacted, and invokes subsequent detection procedures to probe them further. This identification is done on the basis of the nodes that have '0' entries both in the '*From*' and '*Through*' columns in the DRI table. For example, as shown in Table 2, node 7 has not communicated to node 1. Therefore, the node 7 invokes the local anomaly detection procedure for node 1. The 'RTS/CTS' column in the DRI table gives the ratio of the number of *request to send* (RTS) messages to the number of *clear to send* (CTS) messages for the corresponding node. This gives a rough idea about the number of requests arriving at the node for data communication and number packet transmission that the node is actually doing. The significance of the column 'CheckBit' in the DRI table will be discussed in later in this section.

| Node | From | Through | RTS/CTS | CheckBit |
|------|------|---------|---------|----------|
| 1 | 0 | 0 | 15 | 0 |
| 2 | 1 | 1 | 5 | 1 |
| 6 | 0 | 1 | 3 | 0 |
| 8 | 1 | 0 | 6 | 1 |
| 9 | 0 | 1 | 4 | 0 |

Table 2. The DRI table of node 7 as depicted in Fig. 3 (Sen et al., 2007b)

The node that initiates the anomaly detection procedure is called the *initiator node* (IN). The IN first chooses a *cooperative node* (CN) in its neighborhood based on its DRI records and broadcasts a RREQ message to its 1-hop neighbors requesting for a route to the CN. In reply to this RREQ message the IN will receive a number of RREP messages from its neighboring nodes. It will certainly receive a RREP message from the *suspected node* (SN) if the latter is really a grayhole (since the grayholes always send RREP messages but drop data packets probabilistically). After receiving the RREP from the SN, the IN sends a probe packet to the CN through the SN. After the *time to live* (TTL) value of the probe packet is over, the IN enquires the CN whether it has received the probe packet. If the reply to this query is affirmative, (i.e., the probe packet is really received by the CN) then the IN updates its DRI table by making an entry '1' under the column '*CheckBit*' against the node ID of the SN. However, if the probe packet is found to have not reached the CN, the IN increases its level of suspicion about the SN and activates the cooperative anomaly detection procedure, as discussed later in this section.

In Fig. 3, node 7 acts as the IN and initiates the local anomaly detection procedure for the SN (node 1) and chooses node 2 as the CN. Node 2 is the most reliable node for node 7 as both the entries under columns 'From' and 'Through' for node 2 are '1'. Node 7 broadcasts a RREQ message to all its neighbor nodes 1, 2, 6, 8 and 9 requesting them for a route to the CN, i.e., node 2 in the example. After receiving a RREP from the SN (node 1), node 7 sends a probe packet to node 2 via node 1. Node 7 then enquires node 2 whether it has received the probe packet. If node 2 has received the probe packet, node 7 makes an entry '1' under the column '*CheckBit*' in its DRI table corresponding to the row of node 1. If node 2 has not received the probe packet, then node 7 invokes the cooperative anomaly detection procedure. The objective of the cooperative anomaly detection is to increase the detection reliability by reducing the probability of false detection.

The cooperative detection procedure is activated when an IN observes that the probe packet it had sent to the CN through the SN did not reach the CN. The IN invokes the cooperative detection procedure and sends a cooperative detection request message to all the neighbors of the SN. When the neighbors of the SN receive the cooperative detection request message, each of them sends a RREQ message to the SN requesting for a route to the IN. After the SN responds with a RREP message, each of the requesting nodes sends a '*further probe packet*' to the IN along that route. This route will obviously include SN, as SN is a neighbor of each requesting node and the IN as well. Each neighbor of the SN (except the IN) now notifies the IN that a '*further probe packet*' has already been sent to it. This notification message from each neighbor is sent to the IN through routes which do not include the SN. This is necessary to ensure that the SN is not aware about the on-going cross checking process. The IN will receive numerous '*further probe packets*' and notification messages. The IN now constructs a *ProbeCheck* table. The ProbeCheck table has two fields: *NodeID* and *ProbeStatus*. Under the NodeID field, the IN enters the identifiers of the nodes which have sent notification messages to it. An entry of '1' is made under the column 'ProbeStatus' corresponding to the nodes from which the IN has received the 'further probe packet'.

| NodeID | ProbeStatus |
|--------|-------------|
| 2      | 0           |
| 6      | 1           |
| 8      | 1           |
| 9      | 1           |

Table 3. The ProbeCheck table for node 7 (Sen et al., 2007b)

An example ProbeCheck table for node 7 of the network in Fig. 3 is presented in Table 3. It may be observed that node 7 has received the '*further probe packet*' from all the neighbors of the SN (node 1) except node 2. There may be a possibility that the probe packet might have not been maliciously dropped by the SN, rather it has been lost because of collision or buffer overflow. A mathematical estimation can be made for the probability of collision or buffer overflow at the SN (Sen et al., 2007a). However, to avoid complex mathematical computation, we propose a simple mechanism where each node sends three 'further probe packets' interspaced with a small time interval. If none of these three packets from a neighbor are received by the IN, the SN is believed to be behaving like a grayhole for that node during that time. This grayhole behavior may be exhibited for a single node (as for node 2 in Table 3) or may be for a group of nodes.

### 7.5 A Secure and Energy-Efficient Routing Protocol for WSNs

To address the problem of security and efficiency in routing in WSNs, a scheme has been proposed by Sen et al. that reliably identifies compromised (or faulty) nodes and utilizes a routing path that avoids these nodes (Sen et al., 2010). The protocol utilizes a single-path routing concept and thereby saves energy-consumption. The proposed protocol is a modification of the routing scheme proposed in (Lee et al., 2006). However, it is more energy- efficient and less delay-inducing.
The protocol is based on a robust *neighborhood monitoring system* (NMS). NMS works on promiscuous monitoring of the neighborhood by a node and detection of any possible malicious packet dropping attack by a cooperative algorithm using *neighbor list checking* (Sen et al., 2010). The scheme ensures reliable hop-by-hop delivery of packets in a WSN even in presence of malicious nodes that may launch packet-dropping attack in the routing path. To defend against packet-dropping attack, most of the existing algorithms exploit the concept of multi-path routing, where a single packet is routed through multiple paths from the source to the sink. While this approach ensures reliable packet delivery, it consumes an appreciable amount of energy for delivering each packet. To avoid this problem, the protocol uses a single-path routing mechanism. If a malicious node is encountered, the node is avoided and the packet is routed around it in an efficient manner, still in a single-path mode to the base station. The selection of the new path is based on some broadcast signaling in the neighborhood of the malicious node. The salient features of the protocol are briefly described below:

(i) *Neighbor list checking*: during the neighbor discovery phase, each node exchanges *hello* messages with its neighbor nodes to know its 1-hop and 2-hop neighbors (i.e., neighbors of each of its neighboring nodes). The neighborhood information is subsequently verified by exchange of *neighbor list checking* messages (Sen et al., 2010).

(ii) *One-hop packet forwarding*: when a node $u$ sends a packet to its neighbor, it first keeps a copy of the packet in its buffer, and then forwards it to its next-hop node $v$ before encrypting it with the cluster key of the node $u$. Since the cluster key is shared between the node and all its neighbors, the packet encrypted and sent by node $u$ to node $v$ can be overheard by all the neighbors of node $u$.

(iii) *Monitoring nodes selection*: as the packet is forwarded from node $u$ to node $v$, the neighbors of node $u$ that are also neighbors of node $v$ receive the packet and store it in their buffers. These nodes are designated as the secondary monitoring nodes. For example, in Fig. 4, nodes $w$ and $y$ are the secondary monitoring nodes for node $v$. The node $u$ is the primary monitoring node. The nodes that are not neighbors of node $v$ but have received the packet because they are neighbors of node $u$, discard the packet. The primary node knows the secondary monitoring nodes, since every node knows its 1-hop and 2-hop neighbors.

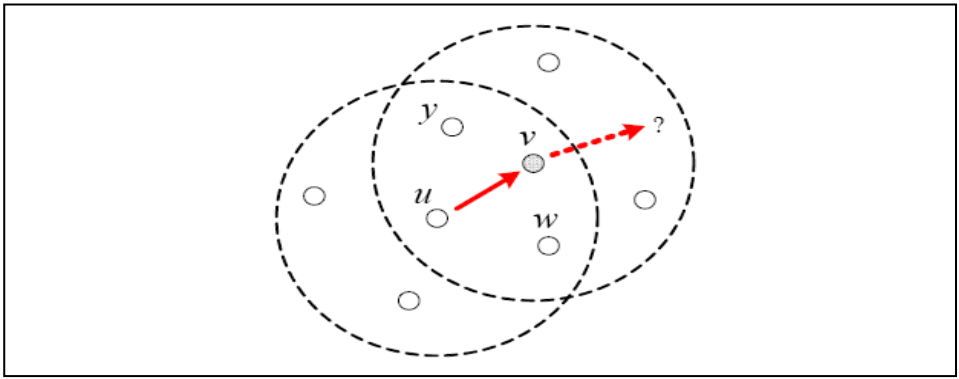

Fig. 4. Neighbor monitor system (secondary nodes $w$, $y$ ; primary node $u$) (Sen et al., 2010)

(iv) *Role of secondary monitoring nodes*: the secondary monitoring nodes $w$ and $y$ monitor the traffic from node $v$ and compare the outbound packets from node $v$ with the packets stored in their buffer. The next-hop address of each packet is also verified to check whether the packet's intended next-hop is a really a neighbor of node $v$, by cross-checking the neighbor list of node $v$. If both these checks yield positive results, the secondary monitoring nodes remove the packet from their buffer and their role of monitoring is complete for that packet. If any packet is found to remain in the buffer of a secondary monitoring node for more than a threshold period of time, it first sends a broadcast signal in its neighborhood to inform all its neighbors that it is going to forward the packet to its designated next-hop so that other neighbors do not forward the same packet. The secondary monitoring node now forwards the packet to its designated next-hop after encrypting the packet with the cluster key. The role of the secondary node now becomes that of the primary node and its neighbors become the secondary node. This is in contrast to the scheme proposed in (Lee et al., 2006), where all the secondary nodes forward the packet in a multi-path mode.

(v) *Role of primary monitoring node*: the role of a primary monitoring node (node $u$) is identical to that of secondary monitoring nodes (nodes $w$ and $y$); the only difference is that it listens not only on the traffic from node $v$, but also on the traffic from the nodes $w$ and $y$. If the packet is correctly forwarded by any one of the nodes $v$, $w$, $y$, the node $u$ removes the packet from its buffer. The role of node $u$ as the primary monitoring node is now complete. If time out occurs for a packet, the primary monitoring node $u$ forwards the packet (encrypted with its cluster key) to its next-hop other than node $v$.

As the packet is routed along a path towards the sink, the above steps of NMS algorithm except the *neighbor list checking* are executed at each hop so that reliable packet delivery can happen through a single path. This is in contrast to the previous schemes proposed in (Ye et al., 2005; Morcos et al., 2005; Yang et al., 2005). In these schemes, a node broadcasts a packet without specifying a designated next-hop, and all neighboring nodes with smaller costs (the cost at a node is the minimum energy required to forward a packet from the node to the base station) or within a specific geographic region continue forwarding the packet to the base station. If nodes $v$, $w$, and $y$ have smaller costs than node $u$ in Fig. 4, then each of them will forward packets received from node $u$ following the existing approaches. However, in the proposed scheme, nodes $w$ and $y$ only observe the packet forwarding activities of node $v$, instead of actively forwarding the packets. In the event of no packet drop, the routing to the base station happens in a single-path, thereby making the process highly energy-efficient. Even in the event of a packet drop, the proposed algorithm works in a single-path mode. This makes it more efficient than the one proposed in (Lee et al., 2006). If the node $v$ in Fig. 4 does not forward the packet it has received from node $u$, then one of the secondary monitoring nodes $w$ and $y$ would forward the packet to its next-hop nodes. The node (either $w$ or $y$) that forwards the packet to its next-hop neighbors will first send a broadcast message in its neighborhood so that its other neighbors would not forward the same packet.

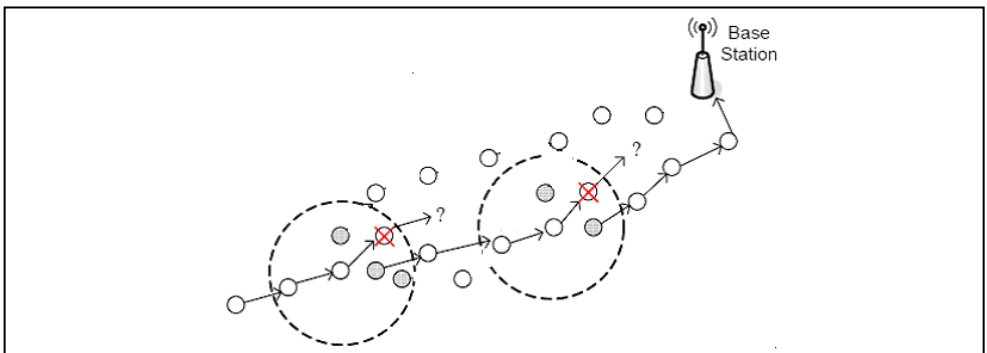

Fig. 5. Two malicious nodes identified by secondary monitoring nodes (Sen et al., 2010)

Fig. 5 shows an example of the application of the scheme, where two malicious (or faulty) nodes are bypassed as the packet is routed to the base station in a single-path.

For the scheme to work, each packet should be encrypted with a cluster key of the forwarding node so that all the neighbors of the forwarding node can decrypt and overhear it. If a link-level encryption was applied between each pair of nodes in the routing path, the scheme would have been more robust, since a compromised node could decrypt only the packets which were destined to it. However, it would have made the scheme less resilient to packet dropping attack. Since encryption with a cluster key provides a reasonable level of robustness to a node compromise, and also supports local broadcast (i.e. resiliency against packet-dropping) it makes the algorithm optimum in its performance (Karlof et al., 2004).

To make the scheme robust to routing disruption attack, where a node intentionally forwards the packets to a spurious address of the next-hop so that the packet is lost in routing, it is necessary that each node should prove that it really has the claimed neighbors. Apparently, a node has the knowledge of its direct neighbors by neighbor discovery and pair-wise key establishment phases discussed earlier. However, in the case of two-hop neighbors, a malicious node $v$ can inform its neighbor $u$ that it also has neighbor node $x$ (any possible id in the network) which in fact is not a neighbor of node $v$ (Fig. 4). Apparently, there is no way node $u$ can detect these false claim of $v$ since $x$ is not in the neighborhood of $u$. To handle this problem, a scheme has been proposed by the authors using which a node can verify the neighbors of each of its neighboring nodes (Sen et al., 2010).

### 7.6 Some More Protocols for Secure Routing in WSNs

Inspired by the work on public key cryptography (Gura et al., 2004; Gaubatz et al., 2004; Watro et al., 2004; Wander et al., 2005), Du et al. have investigated the public key authentication problem (Du et al., 2005). The use of public key cryptography eases many problems in secure routing, for example, authentication and integrity. However, before a node *A* uses the public key from another node *B*, *A* must verify that the public key is actually *B*'s, i.e., *A* must authenticate *B*'s public key; otherwise, man-in-the-middle attacks are possible. In general networks, public key authentication involves a signature verification on a certificate signed by a trusted third party *Certificate Authority* (CA). However, the signature verification operations are very expensive operations for sensor nodes. Du et al. have proposed an efficient alternative that uses only one-way hash function for the public key authentication. The proposed scheme can be divided into two stages. In the pre-distribution stage, A Merkle tree R is constructed with each leaf $L_i$ corresponding to a sensor node. Let $pk_i$ represent node $i$'s public key, $V$ be an internal tree node, and $V_{left}$ and $V_{right}$ be $V$'s two children. The value of an internal tree node is denoted by $\Phi$. The Merkel tree can then be constructed as follows:

$$\Phi(L_i) = h(id_i, pk_i) \text{ for } i = 1,\ldots.N$$

$$\Phi(V) = h(\Phi(V_{left}) || \Phi(V_{right}))$$

In the above expressions, "||" represents the concatenation of two strings and $h$ is a one-way hash function such as MD5 or SHA-1. Let $R$ be the root of the tree. Each sensor node $v$ needs to store the root value $\Phi(R)$ and the sibling node values $\lambda_1, \ldots\ldots \lambda_H$ along the path from $v$ to $R$. If node $A$ wants to authenticate $B$'s public key, $B$ sends its public key $pk$ along with the value of $\lambda_1, \ldots\ldots \lambda_H$ to node $A$. Then, $A$ can use the same procedure to reconstruct the Merkle tree $R`$ and calculate the root value $\Phi(R`)$. $A$ will trust $B$ to be authentic if $\Phi(R`) =$

$\varPhi(R)$. A sensor node only needs $H + 1$ storage units for the extra hash values. Based on this scheme, Du et al. further extended the idea to reduce the height of the Merkle tree to improve the communication overhead of the scheme. The proposed scheme is more efficient than signature verification on certificates. However, the scheme requires that some hash values be distributed in a pre-distribution stage. This results in some scalability issues when new sensors are added to an existing WSN.

Tanachaiwiwat et al. have presented a novel secure routing protocol- *trust routing for location aware sensor networks* (TRANS) (Tanachawiwat et al., 2003). It is primarily meant for use in data centric networks. It makes use of a loose-time synchronization asymmetric cryptographic scheme to ensure message confidentiality. The authors have used *μTESLA* to ensure message authentication and confidentiality. Using *μTESLA*, TRANS is able to ensure that a message is sent along a path of trusted nodes utilizing location aware routing. The base station broadcasts an encrypted message to all its neighbors. Only the trusted neighbors will possess the shared key necessary to decrypt the message. The trusted neighbors then add their locations (for the return trip), encrypt the new message with their shared key, and forward the message to their neighbors closest to the destination. Once the message reaches the destination, the recipient is able to authenticate the source (base station) using the MAC corresponding to the base station. To acknowledge or reply to the message, the destination node can simply forward a return message along the same trusted path from the message was received (Tanachawiwat et al., 2003).

Papadimitratos and Hass have proposed a secure route discovery protocol that guarantees correct topology discovery in an ad hoc sensor network (Papadimitratos et al., 2002). The security relies on the MAC (message authentication code) and an accumulation of the node identities along the route traversed by a message. In this way, a source node discovers the sensor network topology as each node along the route from source to destination appends its identity to the message. In order to ensure that the message has not been tampered with, a MAC is verified at the source and the destination.

A family of configurable secure routing protocols called *secure implicit geographic forwarding* (SIGF) has been proposed in (Wood et al., 2006). SIGF is based on a nondeterministic hybrid routing protocol – IGF (Blum et al., 2003) that is completely stateless. This allows SIGF to handle network dynamics effortlessly, and intrinsically limits the effects of a compromised node to a local area. There are no routing tables to corrupt, since forwarding decisions are made as late as possible – when a packet is ready to transmit over the air. However, the protocol is susceptible to a CTS rushing attack (Hu et al., 2003b).

To defend against route poisoning attack in a multi-hop WSN, a trust-aware routing framework has been proposed in (Zhan et al., 2010). The protocol integrates trustworthiness and energy-efficiency in routing decisions. Each node maintains a neighborhood table with trust level values and energy cost values for certain known neighbors. Once a node is able to decide its next-hop for routing a packet to the base station, it broadcasts its energy-report message that contains the information regarding the energy cost to deliver a packet from the node to the base station. The trustworthiness of a node is computed from its packet forwarding statistics. In this way, a secure and energy-efficient routing is achieved.

Table 4 presents a comparative analysis of some secure routing protocols for WSNs.

Table 4. Comparison of secure routing protocols for WSNs

| Protocols \ Attacks | SPINS (Perrig, '02) | LKHW (Di Pietro, '03) | SecDEACH (Han, '10) | KeyChain (Liu, '03) | LEAP (Zhou, '07b) | SecRoute (Sen, '10) | SIGF (Wood, '06) | TARF (Zhan, '10) | RLEACH (Zhang, '08) | TamperSec (Pecho, '09) |
|---|---|---|---|---|---|---|---|---|---|---|
| Eavesdropping | Yes | Yes | Yes | No | Yes | Yes | Yes | No | Yes | Yes |
| Route poisoning | Yes | Yes | Yes | Yes | No | Yes | Yes | Yes | Yes | Yes |
| Sinkhole/Blackhole | Yes | No | Yes | No | Yes | Yes | Yes | Yes | Yes | Yes |
| Grayhole | Yes | No | Yes | No | Yes | Yes | Yes | Yes | Yes | Yes |
| Wormhole | Yes | No | Yes | No | No | No | No | Yes | Yes | Yes |
| Sybill | Yes | No | No | No | Yes | Yes | Yes | Yes | Yes | Yes |
| Replay | Yes | Yes | Yes | Yes | Yes | Yes | Yes | No | Yes | Yes |
| Hello Flood | Yes | No | Yes | Yes | Yes | Yes | Yes | Yes | Yes | Yes |
| Node impersonation | Yes | Yes | No | No | Yes | Yes | No | Yes | No | Yes |
| Node replication | Yes | Yes | Yes | No | No | Yes | No | No | No | Yes |

## 8. Conclusion

Although research efforts have been made on cryptography, key management, secure routing, secure data aggregation, and intrusion detection in WSNs, there are still some challenges to be addressed. First, the selection of the appropriate cryptographic methods depends on the processing capability of the sensor nodes, indicating that there is no unified solution for all sensor networks. Instead, the security mechanisms are highly application-specific. Second, sensors are characterized by the constraints on energy, computation capability, memory, and communication bandwidth. The design of security services in WSNs must satisfy these constraints. Third, most of the current protocols assume that the sensor nodes and the base stations are stationary. However, there may be situations, such as battlefield environments, where the base station and possibly the sensors need to be mobile. The mobility of the sensor nodes has a great influence on sensor network topology and thus raises many issues in secure routing protocols. Some future trends in WSN security research are identified as follows:

*Exploit the availability of private key operations on sensor nodes*: recent studies on public key cryptography have shown that public key operations are still very expensive to realize in sensor nodes. A public key cryptography can greatly ease the design of security in WSNs, improving the efficiency of private key operations on sensor nodes is highly desirable.

*Secure routing protocols for mobile sensor networks*: mobility of sensor nodes has a great influence on sensor network topology and thus on the routing protocols. Mobility can be at the base station, sensor nodes, or both. Current protocols assume the sensor network is stationary. New secure routing protocols for mobile sensor networks need to be developed.

*Time synchronization issues*: current broadcast authentication schemes such as µTESLA and its extensions require the sensor network to be loosely time synchronized. This requirement is often hard to meet and new techniques that do not have such requirement are in demand.

*Scalability and efficiency in broadcast authentication protocols:* new schemes with higher scalability and efficiency need to be developed for authenticated broadcast protocols. The recent progress on public key cryptography may facilitate the design of authenticated broadcast protocols.

*QoS and security:* performance is generally degraded with the addition of security services in WSNs. Current studies on security in WSNs focus on individual topics such as key management, secure routing, secure data aggregation, and intrusion detection. QoS and security need to be evaluated together in WSNs.